\documentclass[aps,prd,onecolumn,superscriptaddress,showpacs,nofootinbib,floatfix]{revtex4}

\usepackage{dcolumn}
\usepackage{bm}

\usepackage[dvips]{graphicx}
\usepackage{float}
\usepackage{epsfig}
\usepackage{graphicx}
\usepackage{longtable}
\usepackage{rotating}
\usepackage{amsmath}
\usepackage{ulem}
\usepackage{subfigure}
\usepackage{txfonts}

\usepackage{color}

\newcommand{\MSbar}{\overline{\mathrm{MS}}}
\newcommand{\alphas}{\alpha_\mathrm{s}}

\newcommand{\Mf}{m_\mathrm{f}}
\newcommand{\Me}{m_\mathrm{e}}

\newcommand{\Ms}{m_\mathrm{s}}
\newcommand{\Mc}{m_\mathrm{c}}
\newcommand{\Mb}{m_\mathrm{b}}
\newcommand{\Mt}{m_\mathrm{t}}

\newcommand{\Mff}{m_\mathrm{{f\bar{f}}}}
\newcommand{\MV}{m_\mathrm{V}}
\newcommand{\MZ}{m_\mathrm{Z}}
\newcommand{\MW}{m_\mathrm{W}}
\newcommand{\MH}{m_{_{\mathrm{H}}}}
\newcommand{\GF}{G_\mathrm{F}}

\newcommand{\gamgam}{\mathrm{\gamma}\,\mathrm{\gamma}}
\newcommand{\Zgam}{\mathrm{Z}\,\mathrm{\gamma}}

\newcommand{\ffbar}{\mathrm{f\bar{f}}}
\newcommand{\qqbar}{\mathrm{q\bar{q}}}
\newcommand{\ttbar}{\mathrm{t\bar{t}}}
\newcommand{\bbar}{\mathrm{b\bar{b}}}
\newcommand{\ccbar}{\mathrm{c\bar{c}}}
\newcommand{\ssbar}{\mathrm{s\bar{s}}}
\newcommand{\ddbar}{\mathrm{d\bar{d}}}
\newcommand{\uubar}{\mathrm{u\bar{u}}}

\newcommand{\lele}{\ell^{+}\,\ell^{-}}
\newcommand{\elel}{\mathrm{e^+e^-}}
\newcommand{\mumu}{\mathrm{\mu^+\mu^-}}
\newcommand{\tautau}{\mathrm{\tau^+\tau^-}}

\providecommand{\BRprod}{\mathrm{BR}_{\mathrm{prod}}}

\providecommand{\hdecay}{{\sc hdecay}}
\providecommand{\hzha}{{\sc hzha}}
\providecommand{\prophecy}{{\sc prophecy4f}}

\def\bq{\begin{equation}}
\def\eq{\end{equation}}
\def\ba{\begin{eqnarray}}
\def\ea{\end{eqnarray}}

\def\ttt#1{\texttt{\small #1}}

\begin{document}

\title{On the (Gaussian) maximum at a mass $\MH\approx$~125~GeV of the product\\ 
of decay probabilities of the Standard Model Higgs boson}

\author{David~d'Enterria}
\affiliation{CERN, PH Department, 1211 Geneva, Switzerland}

\begin{abstract}
The product of the branching ratios of the standard model (SM) Higgs boson in all decay channels available
below the top-antitop threshold is observed to be a Gaussian distribution of the Higgs boson mass 
with a maximum centered at $\MH\approx$~125~GeV, i.e. exactly at the mass value where a new boson 
has been discovered at the Large Hadron Collider. Such an intriguing observation is seemingly driven by the
different $\MH$-power dependence of the Higgs decay-widths into gauge bosons and fermions with
steep anticorrelated evolutions in the transition region $\MH\in(\MW,2\MW$) below the WW decay
threshold. No other SM Higgs mass value has a better combined signal-strength for the whole set of decay channels. 
Speculative consequences of taking this 
feature as indicative of an underlying
principle, that would force the Higgs mass to be that which maximizes its decay probabilities to
all the SM particles simultaneously, are also discussed as a means to provide possible constraints  
for theoretical extensions of the SM.
\noindent
\end{abstract}


\maketitle


\section{Introduction}

In the standard model (SM) of particle physics, the masses of the electroweak W and Z vector bosons are generated
via the Higgs-Brout-Englert-Guralnik-Hagen-Kibble mechanism~\cite{higgs} which also predicts the existence of 
a new elementary neutral scalar (Higgs) boson. The fermion masses are also generated through Yukawa
interactions with the Higgs field. The mass of the Higgs boson itself, given by the square-root of its
(unknown) self-coupling parameter $\lambda$ 
($\MH^2=2\lambda\,v^2$, where $v$ is the vacuum expectation value of the Higgs field fixed by the Fermi
constant~$\GF$: $v=(\sqrt{2}\GF)^{-1/2}\approx$~246.22~GeV) 
is however a free parameter of the theory.
Theoretically, the value of $\MH$ is constrained from above by requirements of unitarity of the elastic
W$_\mathrm{L}$W$_\mathrm{L}$ cross sections~\cite{WWscatt}, as well as by triviality bounds for large
self-couplings (for large initial values of $\lambda$, the $\log{(Q^2/v^2)}$ running of the self-coupling 
would make it non-perturbative at energies not very far from the electroweak scale)~\cite{Cabibbo:1979ay}. 
Both requirements imply $\MH\lesssim$~700~GeV. 
Lower limits on $\MH$ are imposed by vacuum stability arguments ($\lambda$ cannot be too small since
otherwise the top-quark Yukawa coupling would make it negative)~\cite{Cabibbo:1979ay}. 
The recent experimental observation at the Large Hadron Collider (LHC) of a new boson at
$\MH\approx$~125~GeV in various decay modes consistent with the 
SM Higgs boson~\cite{cms_higgs,atlas_higgs}, constitutes an experimental breakthrough in our understanding of the
mechanism by which all elementary particles acquire their mass.\\ 

Basic conservation rules imply that a neutral spinless SM Higgs boson can only decay into 
(i) pairs of fundamental fermions $\qqbar$ = $\uubar,\ddbar,\ssbar,\ccbar,\bbar,\ttbar$ for quarks, and 
$\lele$ = $\elel,\mumu,\tautau$ for leptons (neglecting neutrinos which are massless in the SM, see later), 
(ii) pairs of (virtual or real) heavy bosons VV~=~ZZ,WW; as well as (iii) pairs of massless bosons
($gg,\gamgam$) and $Z\gamma$ via heavy-particle 
(mostly W and top-quark) loops.
The partial decays widths $\Gamma_i$ of the Higgs boson into the most relevant channels, i.e. the inverse decay
rates to each final state kinematically allowed for a given value of $\MH$, have been determined 
including various high-order quantum chromodynamics (QCD) and electroweak corrections (see~\cite{hdecay} and
refs. therein). The total width is obtained by summing all the Higgs partial widths: $\Gamma_{\mathrm{tot}}=\sum_{i} \Gamma_{i}$.
The branching ratio of a given mode is BR~=~$\Gamma_i/\Gamma_{\mathrm{tot}}$ and, of course, the sum of all of them
amounts to one, $\sum_{i}\,$BR$_{i}$~=~1. 
Decay widths and branching ratios to the various final-states can be computed using programs
such as \hdecay~\cite{hdecay}, and have typical theoretical uncertainties in the range
$\Delta$BR~$\approx$~3--10\% for $\MH\approx$~125~GeV~\cite{Denner:2011mq}. The SM Higgs couplings to
fundamental fermions are proportional to the fermion masses $\Mf$, whereas its couplings to bosons are
proportional to the square of their masses $\MV^2$, a direct indication of the primary role of the Higgs
particle in the breaking of electroweak symmetry. 
The partial $\ffbar$ and VV decay-widths are thus given by the tree-level relations (see e.g.~\cite{Djouadi:2005gi})
\bq
\Gamma_{\ffbar} \;=\;\frac{N_c \, \GF \, \Mf^2}{4\sqrt{2}\,\pi} \cdot \MH \cdot \left(1-\frac{4\Mf^2}{\MH^2}\right)^{3/2},
\;\; {\rm where } \;\; N_c\,=\,3\, (1) \;\, {\rm for \; quarks \; (leptons),}
\label{eq:H_ff}
\eq
\bq
\Gamma_{\mathrm{VV}} \;=\; \frac{\delta_V \, \GF}{16\sqrt{2}\pi} \cdot \MH^3 \cdot \sqrt{1-\frac{4\MV^2}{\MH^2}} \,
                  \left( 1 - \frac{4\MV^2}{\MH^2} + \frac{3}{4}\left(\frac{4\MV^2}{\MH^2}\right)^2 \right)\,,
\;\; {\rm  with } \;\delta_{\mathrm{V}}\,=\,2\,  (1) \, {\rm for \; W \; (Z).}
\label{eq:H_VV}
\eq
From these expressions one sees first that the decays to fermions are basically linear in $\MH$ but follow a $\MH^3$ 
dependence for bosons, i.e. for large enough $\MH$ the decays to W and Z are clearly preferred over fermions.
The loop-mediated $gg$ and $\gamgam,\Zgam$ decay widths (with more complex expressions not reproduced here)  
feature also a cubic dependence on $\MH$, but with a smaller $\Gamma_{\rm loop}$-prefactor than for VV decays. 
Equation~(\ref{eq:H_VV}) is only valid for on-shell vector bosons, i.e. for $\MH\geq 2\MV$. In the region
where one or both W and Z particles can be produced off-shell, the corresponding partial widths are smaller but
rise much faster with the Higgs mass (a power-law fit to $\Gamma_\mathrm{V^{*}V}(\MH)$ in the range
$\MH\approx$~$\MZ$--2$\MW$ computed with \hdecay\ shows a dependence as fast as $\MH^{14}$). 
As a result, the subtreshold WW$^{*}$ and ZZ$^{*}$ branching ratios are larger than those for di-fermions
already for $\MH\approx$~135~GeV and 157~GeV respectively (Fig.~\ref{fig:BRvsMH}).\\

In this work, we investigate whether there is a particular mass region for which the SM Higgs boson has the
maximum number of decay modes effectively available. For this, we construct a variable from the product of
Higgs branching fractions
\bq
\BRprod(\MH)~=~\prod_i\,\mathrm{BR}_{\mathrm{H\to i}}(\MH)\,,
\label{eq:BRsprod}
\eq
and study whether there is a value (if any) of $\MH$ which maximizes such a distribution. 
Maximization of a product of functions are common in (extended) maximum likelihood analyses.
One finds 
surprisingly that the distribution given by Eq.~(\ref{eq:BRsprod}) 
is Gaussian in the range of masses $\MH\approx$~$\MZ$--2$\MW$ with a maximum centered at $\MH\approx$~125~GeV
and a width of $\sigma_{\MH}\approx$~6~GeV. The reasons and possible phenomenological consequences of such an
unexpected behaviour are explored below.

\section{Theoretical setup}

The branching ratios of the Higgs boson decays in the SM are determined following the same
approach developed in Refs.~\cite{Denner:2011mq,Dittmaier:2012vm,Heinemeyer:2013tqa}, i.e. combining the
partial widths computed with \hdecay\ (version 5.11)~\cite{hdecay} and the total Higgs width obtained with \prophecy\ (version 2.0)~\cite{prophecy4f}.
\hdecay\ includes next-to-leading order (NLO) corrections (or higher-order ones in some cases) for the QCD and
electroweak contributions for all decay channels, 
and \prophecy\ includes interference corrections for all decay final-states of the W,Z bosons.
The inclusion of \prophecy\ corrections modifies by less than $\sim$10\% the branching ratios computed with
\hdecay\ alone, mostly in the mass region $\MH$~=~$\MZ$--2$\MW$.
For the input SM parameters of the calculations, the PDG values~\cite{PDG12} listed in Table~\ref{tab:1} are employed.
The strange-quark mass used is the $\MSbar$
value at two different scales $Q$~=~1,\,2~GeV, and the charm and bottom quark masses correspond to the 1-loop
pole values which are less sensitive to uncertainties of the QCD coupling $\alphas$. Running the alternative \hzha\ Higgs decay
code with the same parameters listed here, yields very similar results~\cite{HZHA}.

\begin{table}[hb]
\caption{Input parameters used for the calculations of the Higgs boson branching ratios carried out in this work.
\label{tab:1}}
\begin{center}
\begin{tabular}{lc|lc|lc|lc}\hline
\hspace{0.7mm} $\Ms$ (GeV) & 0.1(2 GeV),0.19(1 GeV) \hspace{0.7mm} & \hspace{0.7mm} $\Me$ (MeV) \hspace{0.7mm} & \hspace{0.7mm}
0.510998928 \hspace{0.7mm} & \hspace{0.7mm} $\MW$ (GeV) \hspace{0.7mm} & \hspace{0.7mm} 80.385 \hspace{0.7mm} &
\hspace{0.7mm} $\alpha(0)$ & \hspace{0.5mm} 1/137.035999074 \\
\hspace{0.7mm} $\Mc$ (GeV) \hspace{0.7mm} & \hspace{0.7mm} 1.28 
\hspace{0.7mm} & \hspace{0.7mm} $m_\mu$ (MeV)
\hspace{0.7mm} & \hspace{0.7mm} 105.658367 \hspace{0.7mm} & \hspace{0.7mm} $\MZ$ (GeV) \hspace{0.7mm} & \hspace{0.7mm}
91.1876 & \hspace{0.7mm} $\GF$ (GeV$^{-2}$) \hspace{0.7mm} & \hspace{0.5mm} $1.1663787\cdot 10^{-5}$ \\ 
\hspace{0.7mm} $\Mb$ (GeV) \hspace{0.7mm} & \hspace{0.7mm} 4.16 
\hspace{0.7mm} & \hspace{0.7mm} $m_\tau$
(MeV) \hspace{0.7mm} & \hspace{0.7mm} 1776.82 \hspace{0.7mm} & \hspace{0.7mm} $\Gamma_\mathrm{W}$ (GeV) \hspace{0.7mm} &
\hspace{0.7mm} 2.085 \hspace{0.7mm} & \hspace{0.7mm} $\alphas(\MZ^2)$ & 0.1184 \\ 
\hspace{0.7mm} $\Mt$ (GeV) \hspace{0.7mm} & \hspace{0.7mm} 173.5 
\hspace{0.7mm} & & & \hspace{0.7mm}
$\Gamma_\mathrm{Z}$ (GeV) \hspace{0.7mm} & \hspace{0.7mm} 2.4952 \hspace{0.7mm} & & \\ \hline
\end{tabular}
\end{center}
\end{table}

The Higgs boson decays into the first-generation fermions ($\uubar$, $\ddbar$, $\elel$) are not directly
implemented in \hdecay\ since, due to their small masses compared to $\MH$, 
they have extremely reduced branching ratios and are of no experimental relevance. From
Eq.~(\ref{eq:H_ff}) the dependence of the first-family branching-fractions on $\MH$ can be obtained 
from the corresponding widths for heavier (e.g. second-generation) fermions via
\bq
\mathrm{BR}_{\uubar,\ddbar}(\MH) \;\approx\;\left(\frac{m_{\mathrm{u},\mathrm{d}}}{\Ms}\right)^2\cdot\mathrm{BR}_{\ssbar}(\MH) \;,\;\; \;\; 
\mathrm{BR}_{\elel}(\MH) \;\approx\;\left(\frac{m_{\elel}}{m_\mumu}\right)^2\cdot\mathrm{BR}_{\mumu}(\MH)\,.
\label{eq:H_1stgen}
\eq
The branching ratios for the quarks of the first generation are obtained using this expression 
with the current mass values $m_{\mathrm{u,d,s}}\approx$~3,\,5,\,100~MeV. Although the resulting $\uubar,\ddbar$ ($\elel$) 
decays are suppressed by factors of order $10^{-3}$ ($10^{-5}$) compared to the $\ssbar$ ($\mumu$) channel, 
the dependence of their BR on $\MH$ is the same as 
for the other heavier fermions. 
Since the light-quark masses are not precisely known, there are
potentially large uncertainties in their associated BR values. As a cross-check we computed the
branching ratios using the two values for the s-quark mass quoted in Table~\ref{tab:1} and found that
although the corresponding absolute BR$_{\uubar,\ddbar,\ssbar}$ values changed by a factor of two, 
the $\MH$-dependence of BR$_{\uubar,\ddbar,\ssbar}(\MH)$ is the same for any Higgs mass value in the range of interest.
Similarly, the same changes of $m_{\mathrm{u,d,s}}$ modify the total Higgs width only at the permille
level given the smallness of these suppressed branching ratios. 
Since we are just interested in the {\it shape} (and possible maxima) of the distribution given by
Eq.~(\ref{eq:BRsprod}), and since uncertainties in BR$_{\mathrm{u,d,s}}$ propagate just as an overall
normalization factor of $\BRprod$$(\MH)$ which is irrelevant for our study, 
the u,d,s mass uncertainties do not play any role in the results presented hereafter.

\section{Results}

\begin{figure}[hbpt!]
\centering
\includegraphics[width=0.85\columnwidth]{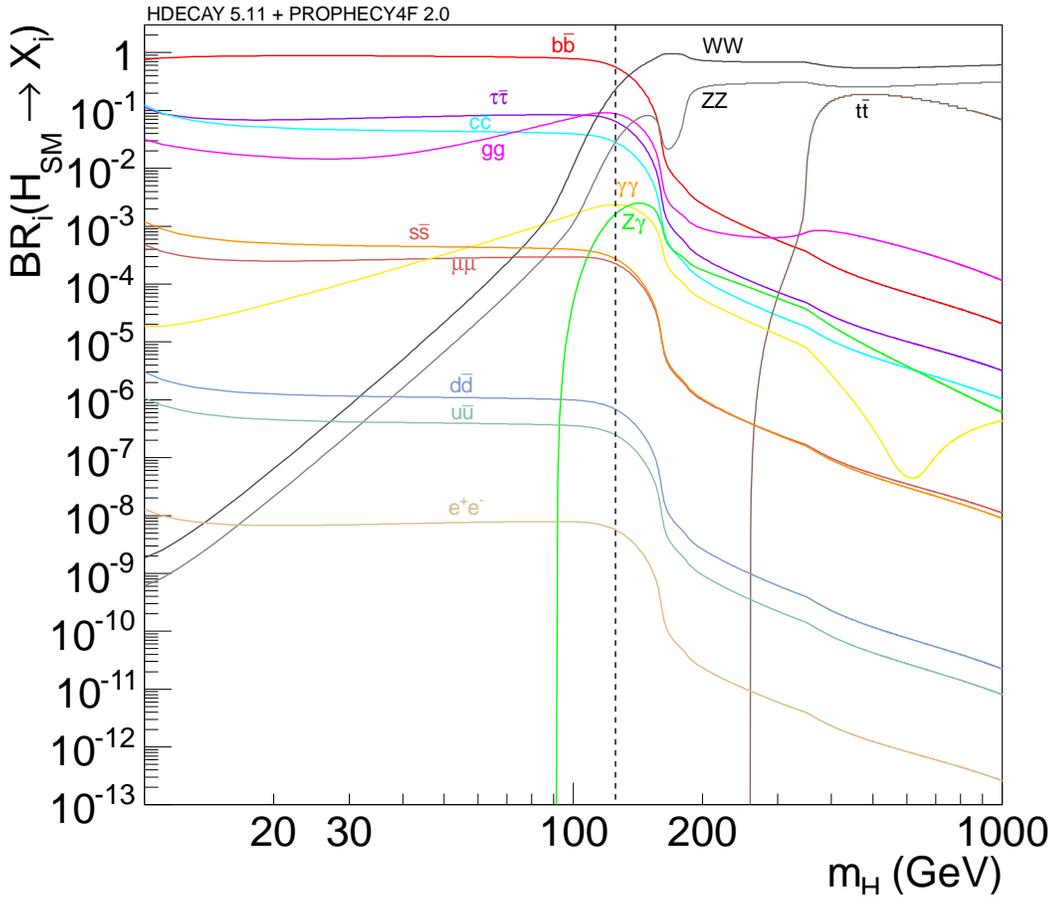}
\caption{Decay branching fractions of the Higgs boson into all fermion and boson pairs of the SM, 
 as a function of the Higgs boson mass, computed with \hdecay\ and
 \prophecy\ with the parameters listed in Table~\ref{tab:1} and using Eq.~(\ref{eq:H_1stgen}) for
 the first-generation fermions. The dashed vertical line indicates
 the position of the 125-GeV boson resonance observed at the LHC~\cite{cms_higgs,atlas_higgs}.}
\label{fig:BRvsMH}
\end{figure}

The branching ratios for all kinematically allowed channels of the SM Higgs boson are computed for masses
$\MH\approx$~0--1~TeV using the setup discussed in the previous Section.
Figure~\ref{fig:BRvsMH} shows the nine fermionic, two bosonic, and three loop-induced BR  
in the mass range $\MH$~=~10--1000~GeV. 
Several observations are worth pointing out. The eight lightest
fermionic channels (plotted here well above their respective thresholds at $\MH=2\Mff$) 
have a relatively flat BR as a function of $\MH$ up to about $\sim$110~GeV, where the very
fast-rising VV$^{*}$ decays start to play an increasing role which subsequently make all the $\ffbar$
branching fractions drop equally steeply. Beyond 2$\MV$ the  bosonic decays are roughly constant as a
function of $\MH$ except for the relatively small modulation due to the onset of the $\ttbar$ mode around
$\MH\approx 2\Mt$, while all other fermionic decays decrease monotonically. The loop-mediated bosonic channels
($gg, \gamgam, \Zgam$) show first a rise with $\MH$ peaking at $\MH\approx$~120--140~GeV --as a matter of fact,
the $\gamgam$ and $gg$ modes seem to peak themselves very close to the Higgs mass value-- followed also by a
monotonic falloff.\\

It is clear from Fig.~\ref{fig:BRvsMH} that all the ``action'' occurs over the mass region ($\MW,\,2\MW$)
centered at around $\MH\approx$~125~GeV (dashed line in Fig.~\ref{fig:BRvsMH}).
The value of the W mass\footnote{Of course, in the SM the Z and W masses are linked through the Weinberg
 mixing angle.}  and the preferred decay of the SM Higgs boson into weak bosons are responsible for all
the features seen in the distribution of branching ratios over this distinctive transition mass region.
Indeed, the fast rise of the bosonic decays, the consequent fast drop of all (light) fermionic modes, 
and the maxima of the loop-induced decays around the region of the newly observed LHC boson resonance, 
are all just driven by the particular value of the W boson mass.\\ 

\begin{figure}[hbtp!]
  \centering
  \subfigure{\includegraphics[width=0.495\columnwidth,height=6.95cm]{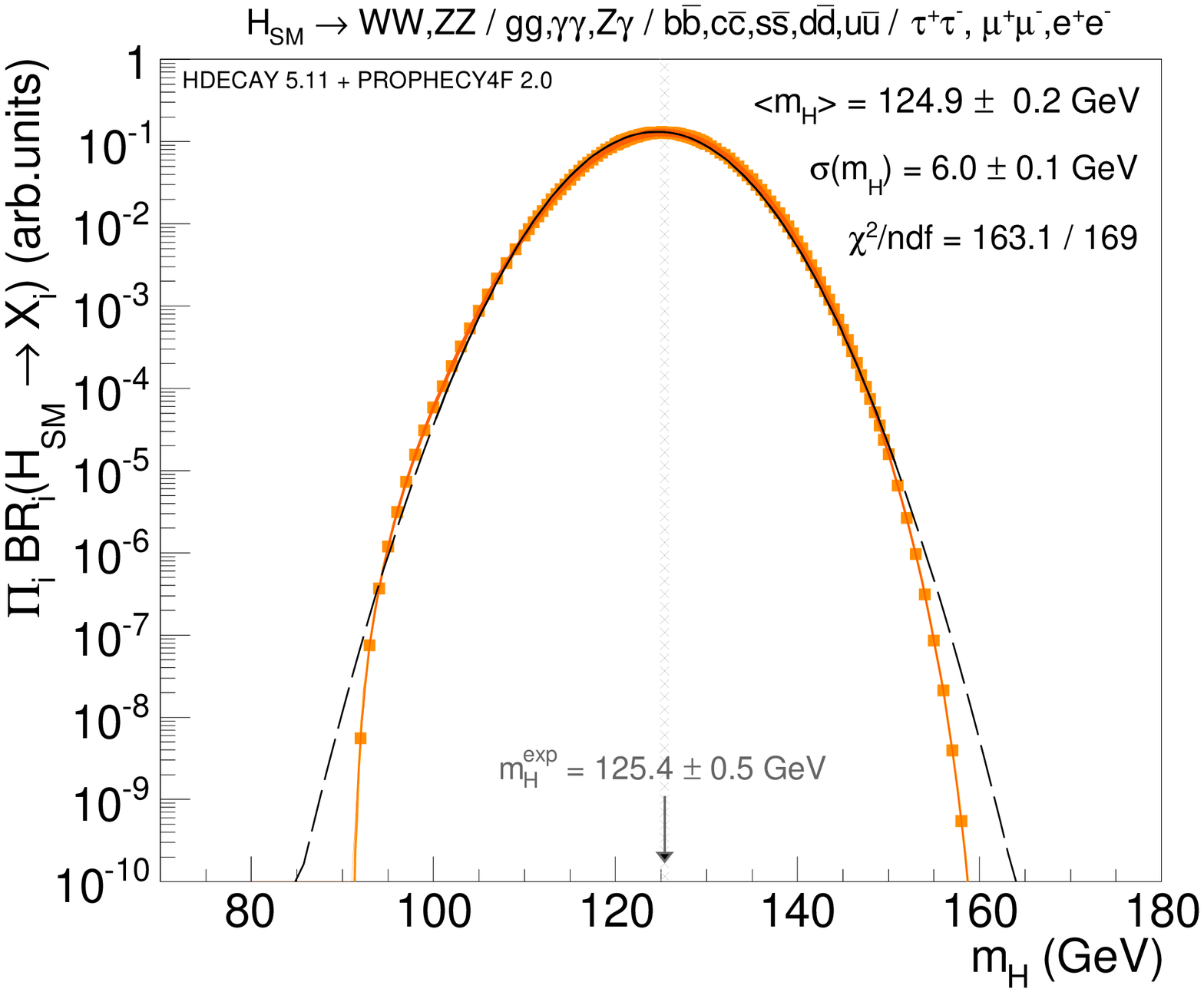}}
  \subfigure{\includegraphics[width=0.495\columnwidth,height=6.9cm]{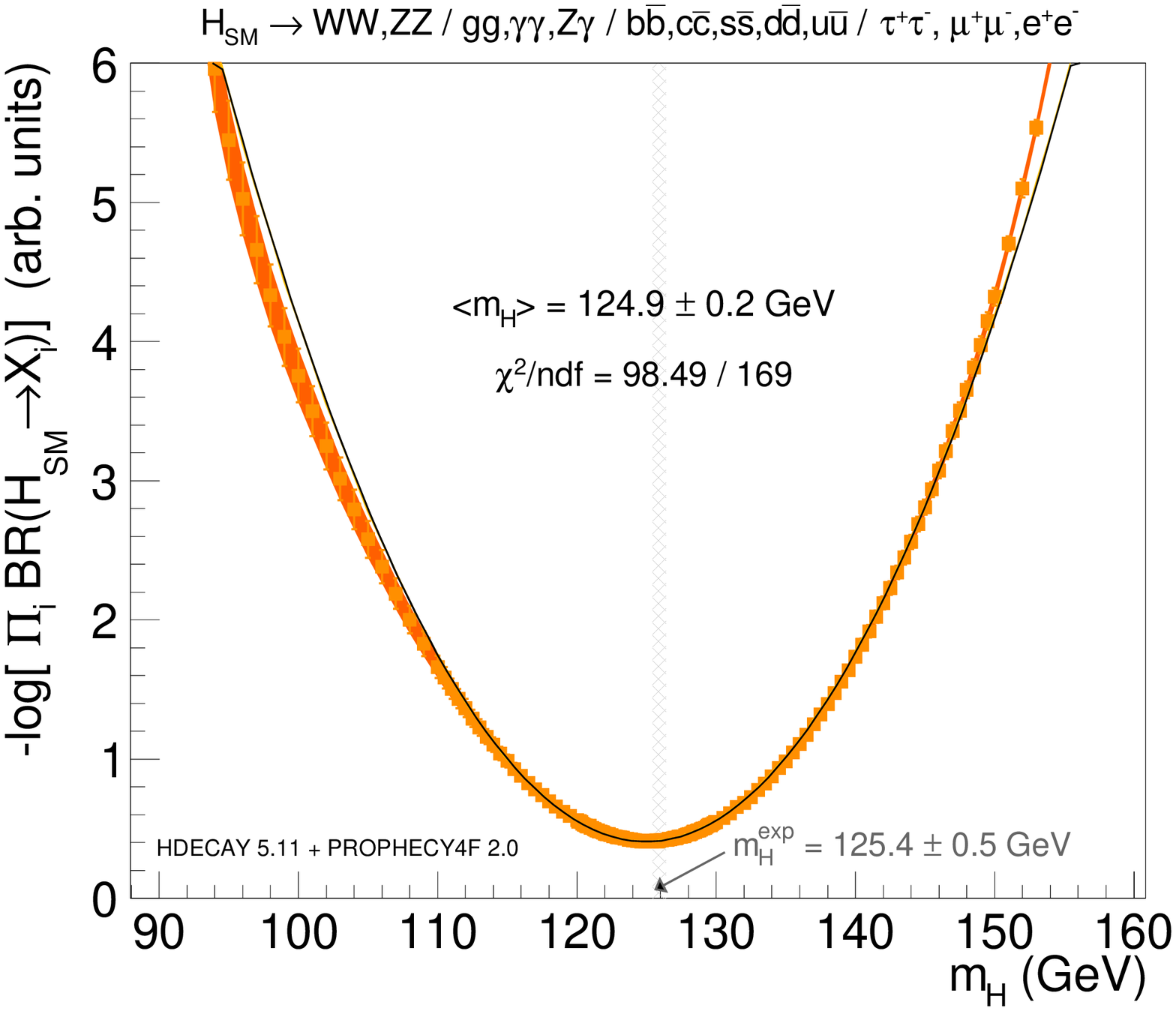}}
  \caption{
   Left: Product of all decay branching ratios of the SM Higgs boson, Eq.~(\ref{eq:BRsprod}), below the 
   $\ttbar$ threshold as a function of the Higgs mass, 
   fitted to a Gaussian distribution (black curve) with the parameters listed in the legend.
   Right: Negative logarithm of the product of Higgs boson branching ratios, $-\log{(\BRprod(\MH))}$ as a
   function of $\MH$, fitted to a quadratic distribution (black curve, whose minimum value is quoted in the figure).
   In both panels, the uncorrelated uncertainties 
   are shown as a band, and the hatched vertical band indicates the combined 
   Higgs boson mass,
   $\MH^{\rm exp}$~=~125.4~$\pm$~0.5~GeV, measured at the LHC~\cite{Chatrchyan:2013lba,ATLAS:2013mma}.}
  \label{fig:BRsprod}
\end{figure}

Figure~\ref{fig:BRsprod} (left) shows the product of all computed branching ratios as a function of $\MH$,
Eq.~(\ref{eq:BRsprod}), below the $\ttbar$ threshold\footnote{The $\ttbar$ branching ratio is actually 
set to zero in \hdecay\ below $\MH\approx$~250~GeV and is omitted in the product of BR($\MH$)
distributions.}. 
The band around the obtained points indicate the corresponding theoretical and parametric
uncertainties, estimated as discussed in~\cite{Denner:2011mq}, 
propagating only those which are uncorrelated across the individual decay channels (this results in
$\BRprod(\MH)$ uncertainties around $\pm$10\% below $\MH\sim$~110~GeV and $\pm$5\% for all other mass values
shown in the plot).
The absolute amplitude of $\BRprod(\MH)$ has no particular meaning and the
distribution has been normalized by its integral.
We note that below $\MZ\approx$~91~GeV, the Z$\gamma$-decay is actually null in \hdecay\ since the code does not include the
Z$^*\gamma$ channel\footnote{Likewise, \hdecay\ does not include the Dalitz $\gamma^*\gamma$ mode with $\gamma^*$
decaying into pairs of fermions --amounting to about 10\% of the $\gamgam$ decay ratio according
to~\cite{Firan:2007tp}-- which nonetheless has the same $\MH$-dependence as BR$_{\gamgam}$ 
and thus does not affect the shape of the obtained $\BRprod(\MH)$ distribution.}. 
This fact explains the sudden drop apparent in the $\BRprod(\MH)$ distribution approaching 90~GeV.

The resulting distribution can be very well fitted by a Gaussian function (black curve), over 6 orders of magnitude in
$\prod_i\,\mathrm{BR}_{\mathrm{H\to i}}(\MH)$, in the transition region from the fermion- to the
boson-dominated regime $\MH$~=~$\MZ$--2$\MW$ with goodness-of-fit per degree-of-freedom: $\chi^2$/ndf~=~163.1/169.
Surprisingly, the distribution has not only a Gaussian shape but it has a maximum centered at 
$\MH =$~124.9~$\pm$~0.3~GeV (where the quoted uncertainty includes the propagated $\BRprod$ uncertainty as well as
systematics variations obtained changing the fit range) and a width of about $\sigma_{\MH}$~=~6~GeV.
The peak of $\BRprod(\MH)$ is coincident to a great accuracy with the mass 
of the Higgs particle discovered at the LHC:
$\MH^{\mathrm{\textsc{cms}}}$~=~125.3~$\pm$~0.4~(stat)~$\pm$~0.5~(syst)~\cite{Chatrchyan:2013lba}, and
$\MH^{\mathrm{\textsc{atlas}}}$~=~125.5~$\pm$~0.2~(stat)~$\pm$~0.6~(syst)~\cite{ATLAS:2013mma},
indicated by a hatched grey band in Fig.~\ref{fig:BRsprod}.
Although the fit quality ($\chi^2/$ndf~$\approx$~1) is very good within 3$\sigma_{\MH}$, 
the product of branching ratios starts to
deviate from a Gaussian in the tail beyond $\MH \approx$~155~GeV. This disagreement may be corrected if one
adds the $\mathrm{t^*\bar{t}^*}$ Higgs decays including the 4-body contributions from diagrams involving two virtual
top-quarks which are very small in amplitude (although steeply rising with $\MH$) and currently neglected in
\hdecay. The Gaussian shape of the branching-ratios product distribution is emphasized in the right panel of
Fig.~\ref{fig:BRsprod} where the (negative) logarithm of the distribution, $-\log{(\BRprod(\MH))}$, 
is plotted and fitted to a quadratic function, whose minimum is at $\MH =$~124.9~$\pm$~0.3~GeV.\\ 

We have also checked what happens 
with $\BRprod$($\MH$) if one excludes one or various decay channels. As a matter of fact, the product of
the two dominant fermion ($\bbar$) and bosonic (WW) decay branching ratios is already a distribution with
a Gaussian-like shape and a maximum at $\MH\approx$~135~GeV with larger width $\sigma_{\MH}\approx$~14~GeV.
The product of the three loop-mediated branching fractions alone is also normally distributed around 
$\MH\approx$~131~GeV with a width $\sigma_{\MH}\approx$~11~GeV.
The consecutive addition of extra fermion-pair BR displaces the Gaussian to lower $\MH$ values and 
reduces its associated width, resulting in the final overall distribution plotted in Fig.~\ref{fig:BRsprod}.\\

The first pragmatical consequence of the quantitative results obtained in Fig.~\ref{fig:BRsprod}
is that a SM Higgs boson at 125~GeV has the largest possible number of decay modes concurrently open for 
experimental study. No other Higgs mass value has theoretically a better signal-strength for the
{\it  whole} set of kinematically-allowed decay channels simultaneously. This fact places the SM Higgs boson at
a mass value of ``maximum opportunity'' to experimentally study its couplings to all the gauge bosons and
fermions at the LHC as well as at any other future collider.\\

The second intriguing observation is the Gaussian shape of the product of Higgs branching ratios.
A priori, there is no physical reason why the product of the many (or a few) branching-ratios distributions 
plotted in Fig.~\ref{fig:BRvsMH} should be normally distributed. However, in statistics theory it is 
well-known that probability densities such as the beta distribution, 
$P(x)\propto x^m(1-x)^n$ for 0~$<~x~<$~1, and the gamma distribution, $P(x) \propto x^m \exp(-x)$ for $x > 0$,
converge to Gaussian distributions\footnote{With means $\mu$~=~$m/(m+n)$, $m$ and widths $\sigma = (m\cdot
  n)/(m+n)^3$, $m$ for the beta and gamma distributions respectively.}
for large values of the powers $m$ and $n$. From the tree-level Higgs-mass dependencies of the decay widths 
given by Eqs.~(\ref{eq:H_ff}) and (\ref{eq:H_VV}):  
$\Gamma_{\ffbar}\propto\MH$, $\Gamma_{\mathrm{VV}}\propto\MH^3$ (and $\Gamma_{\mathrm{loop}}\propto\MH^3$),
the product of branching ratios for the fermionic and bosonic (and loop-mediated) decays 
is indeed a expression which has the following generic parametric form 
\ba
\mathrm{BR}_{\mathrm{prod}}(\MH) & \propto & \bigg(\prod_{i}^{n} \, \Gamma_i(\MH)\bigg) \times \bigg(\sum_i^{n} \Gamma_i(\MH)\bigg)^{-n} 
\propto \left(\MH^{n_{\ffbar}}\cdot \MH^{3(n_{\mathrm{VV}}+n_{\mathrm{loop}})}\right) \times \left(A \cdot n_{\ffbar} \cdot\MH+ B\cdot
  (n_{\mathrm{VV}}+ n_{\mathrm{loop}}) \cdot \MH^{3}\right)^{-n}\,, 
\nonumber
\label{eq:BRsprod_MH}
\ea
for a total number of decay modes $n = n_{\ffbar} + n_{\mathrm{VV}} + n_{\mathrm{loop}}$~=~13 (below the $\ttbar$ threshold) 
and constants $A$ and $B$. Taking $n_{\ffbar}$~=~8, $n_{\mathrm{VV}}$~=~2 and $n_{\mathrm{loop}}$~=~3
one obtains a beta-type distribution with large exponents
\ba
\mathrm{BR}_{\mathrm{prod}}(\MH) & \propto & \MH^{10} \times \left(A' + B' \cdot \MH^2\right)^{-13}
\label{eq:BRsprod_MH}
\ea
which can be indeed a Gaussian-like function for appropriate positive A' and B' constants.\\

The third puzzling observation is the fact that the maximum of the product of BRs peaks to 
within a high accuracy 
at the experimental value of the Higgs boson mass measured at the LHC.
Although the observed Gaussian shape of $\BRprod$($\MH$) might be purely accidental and its maximum
at masses {\it around} $\MH\approx 1.5\MW$ is expected due to the reasons described above, the fact that 
this distribution peaks {\it exactly} at the measured $\MH$ value is completely unexpected.
One can speculate about the possibility that there is an (unknown) underlying dynamical reason that makes the
product of SM Higgs boson branching fractions to peak right at $\MH$. 
Naively, such a possibility is not completely unwarranted if one makes a simplistic analogy with entropy
arguments in the decay of an excited system in equilibrium, to argue that given that the physical realization
of the Higgs field --which gives mass to the rest of elementary particles-- is an unstable particle, the value
of $\MH$ must be such that the Higgs boson has the largest number of ways in which its available mass can be
potentially distributed among the known fundamental particles. 
Any underlying principle that would force the Higgs mass to be that which maximizes its decay probabilities to
all the SM particles simultaneously would, first, imply that the Higgs self-coupling $\lambda$ is not a free
parameter of the theory but it is related to the individual Higgs couplings to the rest of fields.
Should such a mechanism exist, it would require an explanation beyond the SM which could in addition impose
constraints to viable extensions of the theory. For example, fermiophobic or gaugephobic Higgs bosons 
could be already ruled out as they would not lead to the maximum observed in Fig.~\ref{fig:BRsprod}. 
Similarly, extra fermion families with heavier unobserved particles would contribute to the loop-mediated
decays, modifying their $\MH$-dependencies and also, subsequently, changing the shape and possible maximum 
of the $\BRprod(\MH)$ distribution.
In addition, if such a principle existed one could also use it to place limits or constraints on new possible
particles through unobserved Higgs decay channels (such as e.g. to pairs of dark matter particles)
and ``invisible'' decays into e.g. neutrinos and/or axions~\cite{Shrock:1982kd}. 
Adding, for example, three new Higgs boson decays into (Dirac) neutrino-antineutrino pairs with Yukawa
couplings as for the other SM fermions, i.e. using $m_{\nu}\sim\mathcal{O}($meV) instead of $m_{e}$ in
Eq.~(\ref{eq:H_1stgen}), displaces the peak of the distributions shown in Fig.~\ref{fig:BRsprod} to about
$\MH \approx$~122~GeV. It would be interesting to test 
the shape of the corresponding $\BRprod$$(\MH)$ distribution taking into account alternative neutrino mass
realizations e.g. in the context of seesaw mechanisms. Similarly, we defer for a coming study to check whether
the $\MH$-peak in the product of decay fractions present in the SM appears also for the (unconstrained)
masses of other Higgs particles in the context of supersymmetric, or more generally two-Higgs-doublet,
extensions of the SM~\cite{coming}.


\section{Summary}

The fourteen branching ratios of the standard model Higgs boson into the known fermion and boson pairs have
been computed as a function of the Higgs boson mass in the range $\MH\approx$~0--1~TeV with the \hdecay\ and
\prophecy\ programs.
The product of the branching ratios in all decay channels available below the top-antitop threshold 
is observed to be a Gaussian distribution of the Higgs boson mass 
with a maximum centered at $\MH=$~124.9~$\pm$0.2~GeV, i.e. exactly in the region of masses
where a new boson resonance has been discovered at the LHC. We have argued that such an intriguing observation is
in principle not unexpected as the fermionic\footnote{Note that this statement applies to all
fermion pairs, including the top-quark which naively has a much larger scale (2$\Mt$) than $\MW$. 
Indeed, the shape of the $\ttbar$ branching ratio also directly depends on the value of $\MW$ in the region where one
or both top-quarks are off-shell. Also, the top-quark together with the W boson enter indirectly into the
loop-induced decays with massless gauge bosons in the final state and determine their behaviour (local maxima) 
in the region $\MH\in$~($\MW$,$2\MW$).}, bosonic and loop-mediated Higgs decay branching ratios feature all 
strongly anti-correlated changes around $\MH\approx 1.5\MW$ due to their different $\MH$-power dependencies,
and the product of their branching ratios has a beta-type probability distribution which
converges into a Gaussian-shape for large number of decay channels around this mass region.\\

On the other hand, the observation that the product of Higgs decay probabilities peaks {\it exactly} 
at $\MH$ is more puzzling. If such a fact is not accidental but indicative of 
some (unknown) underlying physical principle 
--i.e. if there is a fundamental mechanism dictating that the value of $\MH$ must be such that it has the
largest number of ways in which the available mass can be potentially distributed into the known fundamental
particles-- it would certainly have consequences and provide hints for constructing extensions of the
SM. First, it would imply that the Higgs self-coupling $\lambda$ is not a free parameter of the theory but it
is directly connected to the individual Higgs couplings to the rest of fields. Second, such a 
principle could already be used to rule out e.g. fermiophobic or gaugephobic Higgs bosons as well as extra fermion
families. Last but not least, it could be exploited to impose extra constraints on new hypothesized particles
through their associated Higgs decay channels.\\

In any case, and beyond any speculation, the observation reported here demonstrates quantitatively that the 
SM Higgs boson occurs at a mass value of ``maximum opportunity'' in terms of the experimental study of its
couplings to all the gauge bosons and fermions.  No other SM Higgs mass value has theoretically a better ``signal
strength'' for the {\it  whole} set of kinematically-allowed decay channels simultaneously.
Such a fact will help, in practical terms, to maximize the number of constraints that can be imposed to any
physics beyond the standard Model at the LHC and at any other future collider.


\section*{Acknowledgments}

\noindent
Valuable discussions with Andr\'e David, Albert de Roeck, Michelangelo Mangano and Patrick Janot 
are acknowledged. 
I gratefully acknowledge partial support of the Perimeter Institute for Theoretical Physics and of
Brookhaven National Lab where part of this work was undertaken.
                    


\end{document}